\documentclass[%
 apl,%
 amsmath,amssymb,%
 reprint,%
 groupedaddress,
]{revtex4-1}

\usepackage{graphicx}% Include figure files
\usepackage{float}

\begin{document}

\preprint{AIP/123-QED}

\title{Polarimetric analysis of stress anisotropy in nanomechanical silicon nitride resonators}

\author{T. Capelle}
 \altaffiliation[Also at ]{Laboratoire Kastler Brossel, Paris.}%Lines break automatically or can be forced with \\
\author{Y. Tsaturyan}%
\author{A. Barg}
\author{A. Schliesser}
\email[E-mail: ]{albert.schliesser@nbi.dk}
\affiliation{ 
Niels Bohr Institute, Blegdamsvej 17,
2100 Copenhagen, Denmark}

\date{\today}% It is always \today, today,

\begin{abstract}
We realise a circular gray-field polariscope to image stress-induced birefringence in thin (sub-micron thick) silicon nitride (SiN) membranes and strings.
This enables quantitative mapping of the orientation of principal stresses and stress anisotropy, complementary to, and in agreement with, finite element modeling (FEM).
Furthermore, using a sample with a well known stress anisotropy, we extract a new value for the photoelastic  (Brewster) coefficient of silicon nitride, $C \approx (3.4~\pm~0.1)\times~10^{-6}~\mathrm{MPa}^{-1}$.
We explore possible applications of the method to analyse and quality-control stressed membranes with phononic crystal patterns.
%Valid PACS numbers may be entered using the \verb+\pacs{#1}+ command.
\end{abstract}

\maketitle

Silicon nitride membranes and strings under high tensile stress have excellent mechanical and optical properties \cite{Zwickl2008, Wilson2009}, making them a widely used platform to study the behaviour of mechanical systems in the quantum regime \cite{Purdy2013,Purdy2013a, Zhou2013,Underwood2015, Wilson2015, Peterson2016, Nielsen2016, Koji2016}.
Recently, further enhancement of these properties through in-plane patterning has been explored.
Examples include one-\cite{Stambaugh2014} and two-dimensional\cite{Chen2016a,Norte2016,Bernard2016} subwavelength optical grating and photonic crystal structures which can boost the reflectivity of SiN beyond $99.9\%$.
In the mechanical domain, ``trampoline'' resonators  \cite{Kleckner2011} combine thin tethers with a light central pad to achieve low-mass, low-frequency resonators with high quality factors \cite{Norte2016, Reinhardt2016, Weaver2016}.
Patterning with phononic bandgap structures \cite{Ghadimi2016,Barasheed2015,Tsaturyan2016} can suppress radiation losses and, if combined with optimised dissipation dilution through ``soft clamping''\cite{Tsaturyan2016}, yield extreme quality factors and room-temperature $Q\cdot f$-products beyond $10^{14}\,\mathrm{Hz}$.

In all instances of patterning, the stress relaxes to a new equilibrium distribution according to the pattern boundary conditions.
This changes dramatically the mechanical, and potentially, via photoelastic coupling, optical properties of the structure.
In absence of a laboratory diagnostic instrument, it has so far been necessary to rely on FEM to simulate the stress redistribution.
In addition, little is known about the photoelastic coupling in silicon nitride  \cite{Borkje2012,Grutter2015}.
To address these deficiencies, we have realised a highly sensitive polarimetric setup which allows quantitative imaging of stress-induced birefringence.
%

%\section{Brewster coefficient extraction}

\begin{figure*}[bth]
\includegraphics[width=\linewidth]{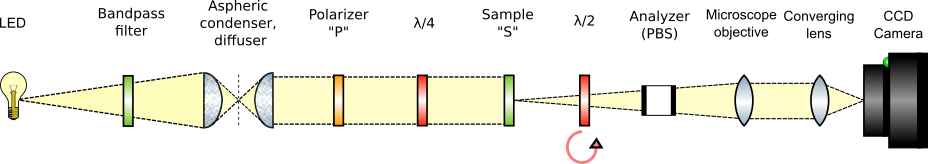}% Here is how to import EPS art
\caption{\label{fig:setup} Gray field polariscope setup. A stepper motor rotates the half wave plate. See text for more details.}
\end{figure*}

Among the numerous %, well-known
possibilities to implement an imaging polarimeter (or polariscope)\cite{Ajovalasit2015}, we have chosen to build a circular gray-field polariscope.
Its basic idea is to illuminate the sample with circularly polarised light, and analyse the ellipticity of the transmitted beam's polarisation in a spatially resolved manner.
This approach has a decisive advantage over plane polariscopes working with linear polarisation when it comes to measuring small optical retardation $\delta$: as we will demonstrate, in the circular polariscope the optical signal is $\propto \delta$, whereas in the plane polariscope it is only $\propto \delta^2$.
In our setup (Fig.~\ref{fig:setup} and Tab.~\ref{tab:setup_specs}), we use a light emitting diode (Thorlabs M780LP1) as a light source, followed by a bandpass filter that eases the requirements on achromaticity of the subsequent optical elements. 
The source is imaged on a diffuser which, together with an aspheric condenser lens, provides a K\"ohler-like illumination of the sample.
Before reaching the sample, the circular polarisation state is defined by a high-contrast polariser and a quarter-wave plate($\lambda/4$).
After the sample, a motor-controlled rotating half-waveplate and a polarising beam splitter cube analyse the polarisation state.
Rotating the waveplate was found to yield better results than rotating various kinds of polarisers, which tended to displace the beam and thus create image artifacts.
A microscope objective (Achrovid 5x) and a plano-convex lens image part of the sample on a CCD camera (Mightex CGE B013-U) with a magnification of $10$.

% serves as the illumination of a K
% Specifically, we use a 
% In order to address those two topics, we set up a gray field polariscope : circularly polarized light illuminates the sample, and an analyzer, composed of a half waveplate whose principal axis can be rotated using a stepper motor, and a fixed linear polarizer, allow to project the output polarization on an arbitrary axis, allowing to measure the stress induced birefringence in the sample. The setup is presented in figure \ref{fig:setup}, and technical characteristics are listed in table \ref{setup_specs}. The choice of an incoherent light source was dictated by the strong diffraction pattern expected in such small features, whereas the choice of an analyzer composed of a rotating half waveplate associated with a fixed linear polarizer instead of a rotating linear polarizer, which corresponds to the usual gray field polariscope, was forced by the fact that linear polarizers slightly deviate the beam, which in a rotation induces a rotating translation of the whole image in the CCD. Whereas a lot of different setups are possible to analyze birefringence in a material using polarimetry, our choice was constrained by the very small signal we want to measure. Indeed, plane polariscopes are for instance sensitive to the square of the stress induced retardation, and circular polariscopes are more sensitive to quarter waveplate errors, which can be strong for an incoherent light source. All the specificities of  the various polariscope setups are well known in the material science community \cite{Ajovalasit2015}.

\begin{table}[tbh]
\begin{center}
\begin{tabular}{c c}
\hline
LED central wavelength (nm)& 780\\
Bandpass filter bandwidth (nm) & 10\\

Working distance of the microscope objective (mm) & 37\\

Focal length of the imaging lens (mm)& 400\\

Pixel size ($\mu$m) & 3.75$\times$3.75\\

Resolution of the CCD camera & 1280$\times$960\\

Bit depth of the CCD camera ADC & 12\\
\hline
\end{tabular}
\caption{Technical characteristics of the polarimetric setup.}
\label{tab:setup_specs}
\end{center}
\end{table}

It is straightforward to compute the expected signal via Jones calculus.
Each area element of the sample can be treated as a general retarder described by the Jones matrix 
\begin{equation}
\label{general retarder}
\textbf{S}_{\delta,\theta}=\textbf{R}^{-1}_{\theta}\cdot\begin{pmatrix}
e^{-i\delta/2}&0\\
0&e^{+i\delta/2}
\end{pmatrix}\cdot\textbf{R}_{\theta},
\end{equation}
where $\theta$ is the azimuthal angle of the polarisation eigenstate basis with respect to a fixed laboratory reference, $\delta$ is the retardation phase, and 
\begin{equation}
\label{rotation matrix}
\textbf{R}_{\theta}=\begin{pmatrix}
\cos(\theta)& -\sin(\theta)\\
\sin(\theta)& \cos(\theta)
\end{pmatrix}
\end{equation}
is the canonical rotation matrix.
Specifically, in the case of stress birefringence of a SiN membrane, $\theta$ denotes the direction of the principal stress in the membrane plane, and $\delta$ is the retardation induced by stress anisotropy,
\begin{equation}
\label{Brewster relation}
\delta=\frac{2\pi}{\lambda}\,C\,l \cdot \Delta\sigma,
\end{equation}
where $\lambda$ is the wavelength of the light source, $\Delta\sigma$ the stress anisotropy, $l$ the thickness of the sample and $C$ the Brewster coefficient, which is a material constant.
The other waveplates can easily be represented by similar matrices, setting $\delta = \pi/2$ for quarter waveplates, and $\delta=\pi$ for half waveplates.
The analyzer at the end projects the polarisation state, as described by the matrix 
\begin{equation}
\label{polarizer}
\textbf{P}_0=\begin{pmatrix}
1 & 0\\
0 & 0
\end{pmatrix}.
\end{equation}
If we introduce $\alpha$ as the angle of the output half waveplate with respect to the output polariser, the intensity at each camera pixel is given by $I_{\mathrm{gray}}\propto\vec{J}_{\mathrm{out}}^{\dagger}\cdot\vec{J}_{\mathrm{out}}$, with
\begin{equation}
\label{}
\vec{J}_{\mathrm{out}}=\textbf{P}_{0}\cdot\textbf{S}_{\pi,\alpha}\cdot\textbf{S}_{\delta,\theta}\cdot\vec{J}_{\mathrm{in}},
\end{equation}
assuming, for simplicity, a perfect input polarisation ${\vec{J}_\mathrm{in}=(1,-i)^{T}/\sqrt{2}}$.

We denote with $\tilde{I}_{\mathrm{gray}}$ the Fourier transform of the output signal with respect to time $t$, during which the half-wave plate rotates with constant angular velocity $\omega \equiv \frac{\partial \alpha}{\partial t}$.
Then it is easy to show that
\begin{equation}
\label{output signal fourier}
\frac{\tilde{I}_{\mathrm{gray}}(4\omega)}{\tilde{I}_{\mathrm{gray}}(0)}=\frac{1}{2}ie^{-2i\theta}\sin(\delta).
\end{equation}
We can thus extract the entities of interest by calculating
\begin{equation}
\label{retardation extraction}
\delta=\arcsin\left(2\left\lvert\frac{\tilde{I}_{\mathrm{gray}}(4\omega)}{\tilde{I}_{\mathrm{gray}}(0)}\right\rvert\right)
\end{equation}
and
\begin{equation}
\label{angle extraction}
\theta  = -\frac{1}{2}\arg\left(\frac{\tilde{I}_{\mathrm{gray}}(4\omega)}{\tilde{I}_{\mathrm{gray}}(0)}\right)+\theta_{0}.
\end{equation}
Here, $\theta_0$ is an offset angle, given by the angle of the first polariser plus $\pi/4$, which corresponds to the position of the quarter waveplate.
Indeed, there is no phase shift when the axis of the quarter waveplate is aligned with the principal axis of the sample.

The data are acquired and processed as follows: images are taken at 10 angular positions for one full rotation.
At each position we acquire a large number of frames (typically 150), from which a frame acquired with the shutter closed is subtracted. This is done in order to remove the dark current, which corresponds to an offset due to the thermal electrons detected by the CCD camera.
The procedure is repeated without the sample to acquire background images.
For each image, treated as a two-dimensional matrix, we calculate the normalized quantity $\tilde{I}_{\mathrm{gray}}(4\omega)/\tilde{I}_{\mathrm{gray}}(0)$. Finally, the background is subtracted from the measurements involving the sample.
This yields a complex matrix, which is translated into the orientation $\theta$ and retardation $\delta$ following eqs.~(\ref{retardation extraction})~and~(\ref{angle extraction}).
%

%\section{Brewster coefficient extraction}
% Are there sections in APL papers?

For initial validation, we performed a measurement on a sample with a particularly simple geometry, namely a $210$~nm thick  silicon nitride ribbon.
We used standard nanofabrication techniques to realise this sample, starting with low-pressure chemical vapor deposition (LPCVD) of $210$~nm stoichiometric silicon nitride on a $500~\mu$m double-side polished silicon wafer.
The chosen deposition parameters create a film with an isotropic tensile stress of ca.~$1{.}2~\mathrm{GPa}$.
Subsequently, ribbons are defined by UV photolithography and reactive ion etching.
The photoresist is removed and the wafer is stripped of its native silicon oxide layer by a buffered hydrofluoric acid (BHF) dip.
Finally, the ribbons are released using an anisotropic KOH etch from the wafer's backside.
Since the ribbons are clamped only from two sides, the stress in the free direction relaxes close to zero.

\begin{figure}
\includegraphics[width = 0.5\textwidth]{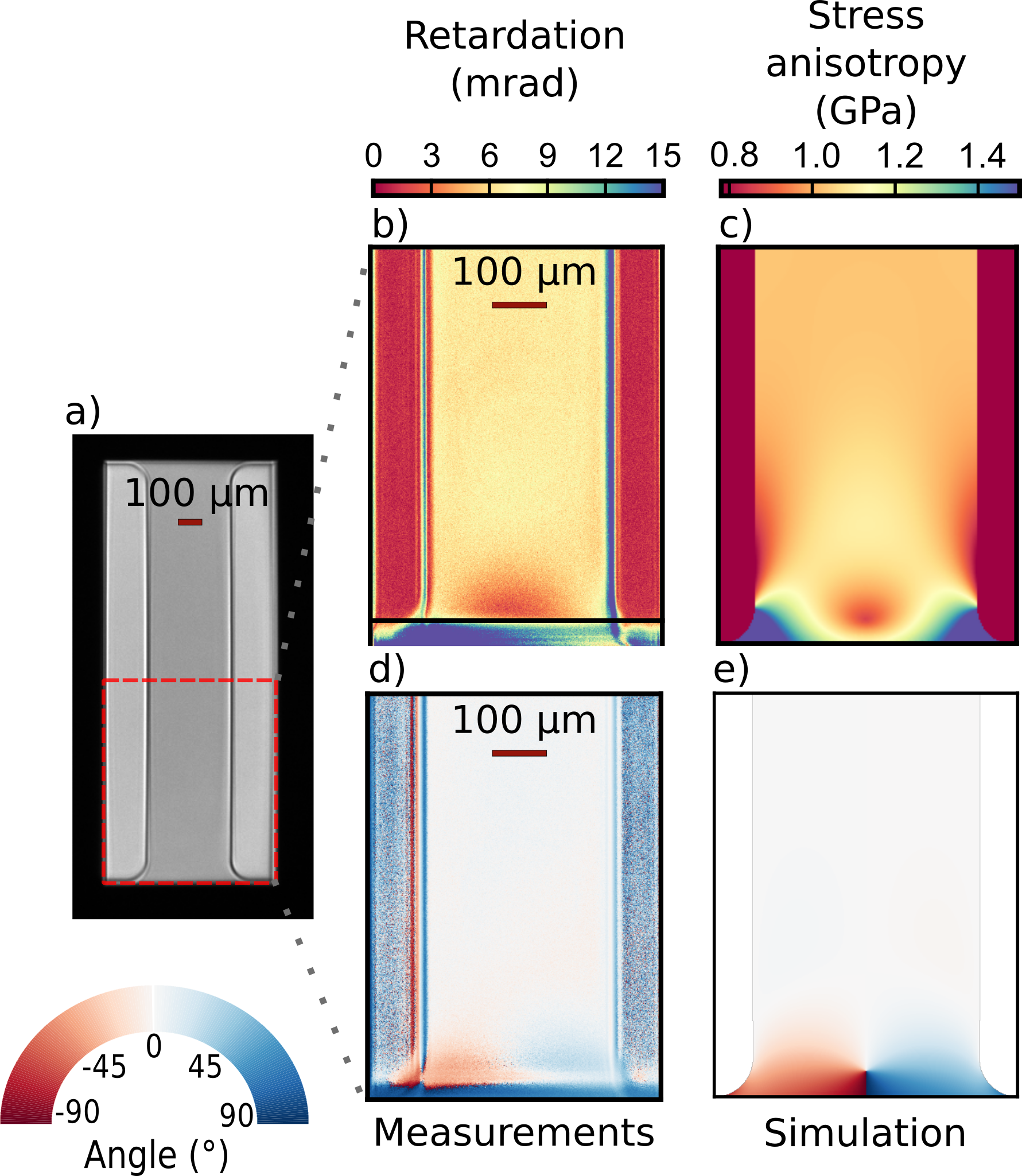}
\caption{\label{ribbon} Analysis of a SiN ribbon. a) CCD photograph of the sample, ribbon shows dark gray. b) Measured retardation $\delta$ c) Simulated stress anisotropy $\Delta\sigma$ d) Measured angle $\theta$ of the optical axis e) Simulated angle of principal stress.}
\end{figure}

Figure \ref{ribbon} shows the measured stress birefringence of such ribbons as obtained using our method. 
The results agree well with our physical intuition, as well as finite-element modelling (COMSOL Multiphysics):
at the center of the ribbon the stress is large along the direction of the ribbon and nonexistent in the orthogonal direction, so the anisotropy is maximal.
Closer to the clamping region the ribbon also gets stressed in the orthogonal direction, due to the proximity to the silicon support, reducing the anisotropy.
The higher stress anisotropy in the ribbon's fillets could not be recovered in our setup, possibly due to diffraction artifacts (see below).
The stress direction is also reproduced correctly: homogeneously along the ribbon in its center, while close to the clamp it rotates by $\pi/4$, preserving the symmetry of the structure.

\begin{figure}
\includegraphics[width = 0.4\textwidth]{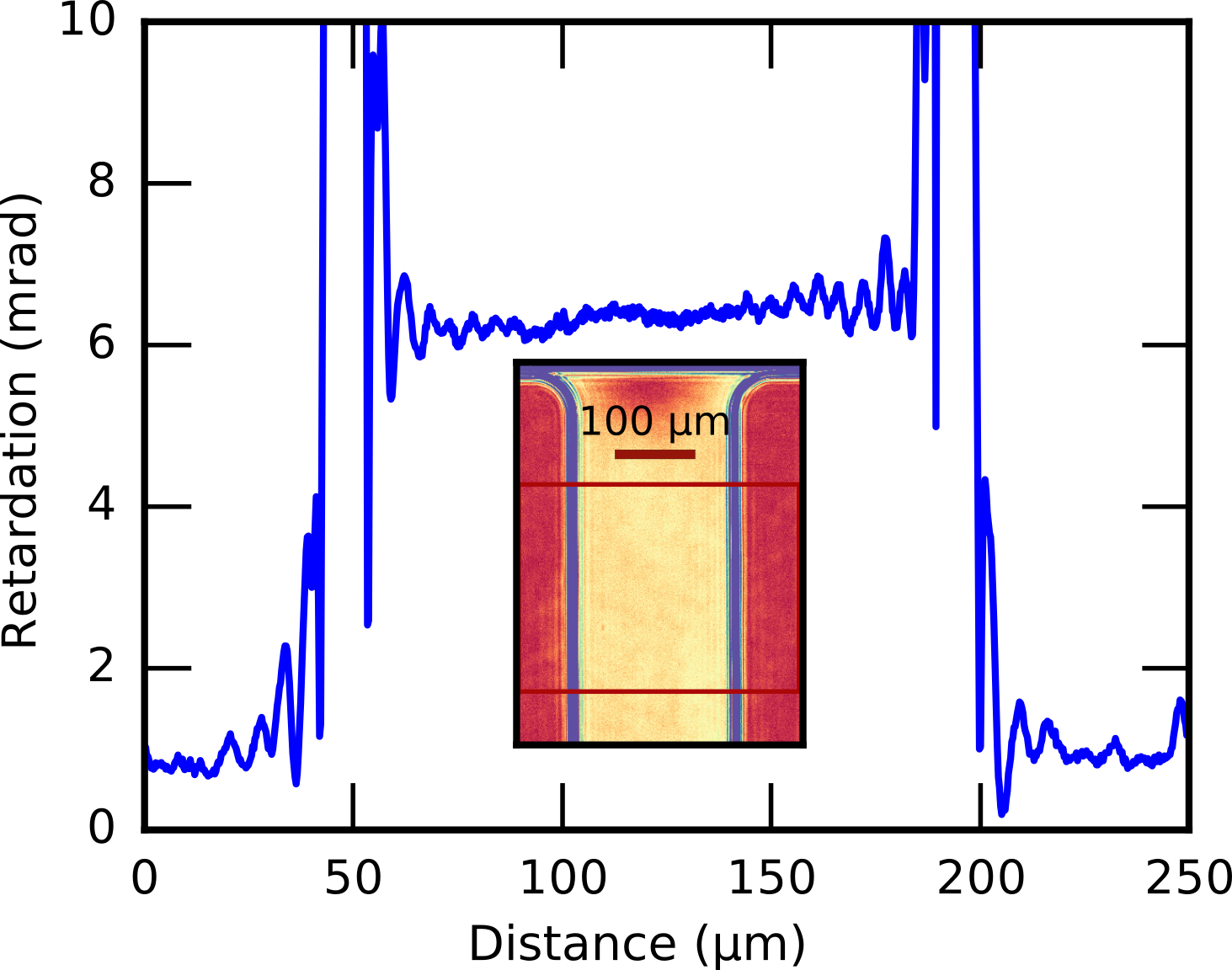}
\caption{\label{zoom} Retardation measurement on a sample with calibrated stress. Shown is an average over all horizontal line cuts between the two red lines shown in the inset, where stress is approximately homogeneous. Inset shows full retardation image.}
\end{figure}

The good agreement prompts us to apply this approach to measure the Brewster coefficient, relating stress anisotropy with birefringence, of SiN, for which quantitative data is  scarce \cite{Borkje2012,Grutter2015}.
To that end we measured a second ribbon, whose initial stress was carefully determined to be $\sigma = 1190$ MPa before releasing of the ribbon\footnote{This measurement was conducted on a separate sample with a $238.6$-nm thick SiN layer grown in the same conditions as the ribbon.}. In order to take into account the change of the boundary conditions during the release of the ribbon, we have to correct this value by a factor $1-\nu$, where $\nu$ is the Poisson ratio of the material, assumed here to be equal to $0.27$.
The result is presented in Figure \ref{zoom}.
We obtain a retardation of $6.4\pm 0.2~\mathrm{mrad}$ by averaging a $\sim 170\times260~\mathrm{\mu m}^2$-area in the central region of the ribbon.
For a quantitative comparison, we correct this value for the multiple reflections inside the film, which are not entirely negligible due to the relatively high ($n\approx 2{.}0$) refractive index of SiN.
To do so, we  used a transfer matrix model for the complex transmittance of this sample\footnote{the transmittance of a membrane of thickness $d$ and refractive index $n$ is given by $t = \frac{2ine^{-ikd}}{2in\cos(kdn)+(1+n^2)\sin(nkd)}$}, expanding it to first order in a small variation of refractive index.
This yielded a correction parameter of $\eta = 1.26$ for the effective thickness of the sample.
The Brewster coefficient of SiN can then be evaluated as
\begin{equation}
\label{brewster coefficient}
C = \frac{\delta}{2\pi}\frac{\lambda}{\eta l}\frac{1}{\sigma(1-\nu)} \approx (3.4~\pm~0.1)\times~10^{-6}~\mathrm{MPa}^{-1},
\end{equation}
where $l=210$ nm is the sample thickness. 
Remarkably, this is two orders of magnitude lower than the value proposed by Campillo \textit{et al.} \cite{Campillo2002}, but close to the value of silica ($C\approx 4\times 10^{-6}~\mathrm{MPa}^{-1}$) \cite{Priestley}, another amorphous transparent dielectric. This mistake arises from the fact that this previous work measured the refractive index and the stress for different $\mathrm{SiH}_2\mathrm{Cl}_2\mathrm{:NH}_3$ gas flow ratios during the LPCVD process, and not the variation of the refractive index due to mechanical stress.

%\section{Application to nanoresonators}

With such calibration at hand, we proceed to applying polarimetric stress analysis to more complicated resonator structures.
Membrane resonators with phononic bandgap shield are of particular interest.
In the context of silicon nitride membranes it has been shown that phononic crystal structures can suppress transmission of vibrations \cite{Tsaturyan2014,Yu2014},  resulting in suppression of dissipation through phonon tunneling \cite{Yu2014,Wilson-Rae2008}, whose avoidance had previously required delicate and often unreliable clamping techniques\cite{Wilson2009}.
Patterning a phononic crystal structure directly onto the membrane not only suppresses phonon tunneling losses, but also enhances the dilution of internal losses dramatically, enabling an increase in the quality factor by more than an order of magnitude \cite{Tsaturyan2016}, as compared to conventional membrane resonators embedded in silicon phononic crystal structures. 
Stress analysis using polarimetry in such complex geometries can be a reliable and simple tool to assess the periodicity of stress anisotropy as required for the formation of a bandgap. 

\begin{figure}
\includegraphics[width = \linewidth]{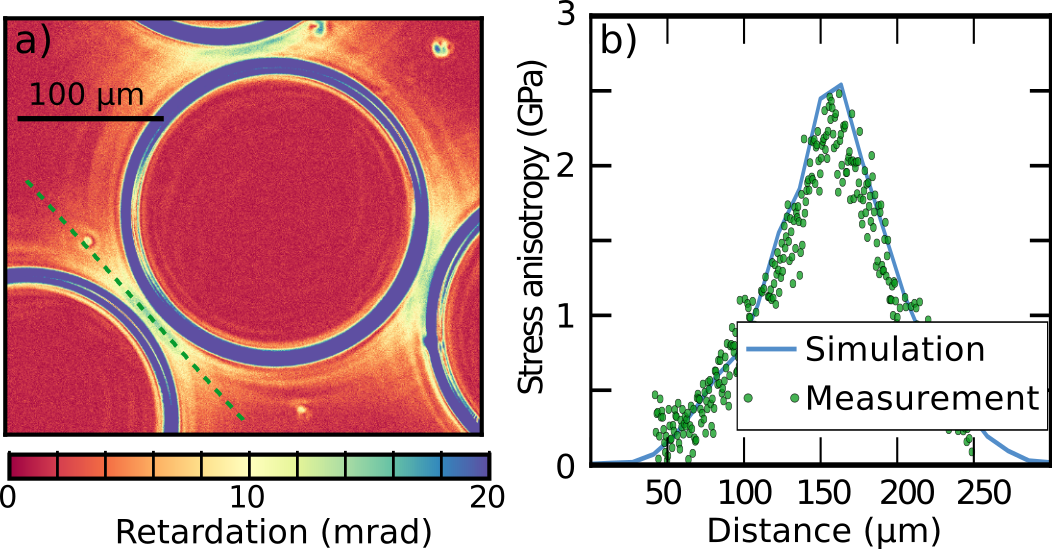}
\caption{\label{patterned_membrane} Analysis of a patterned membrane. a) Retardation image.  b) Measured retardation along the dashed line in a), following a tether between two holes. Values are converted to stress anisotropy via the previously extracted Brewster coefficient, and corrected for multiple reflections in the film.}
\end{figure}

To demonstrate this potential, we realised  membranes (${l\approx 210~\mathrm{nm}}$) with a honeycomb pattern of ${\sim200~\mathrm{\mu m}}$-diameter holes, fabricated using the same techniques as for the ribbons.
Such a patterned membrane exhibits a bandgap as previously described\cite{Tsaturyan2016}.
Once again, we performed a measurement of the retardation, as described above, with the result shown in Figure \ref{patterned_membrane}. 
As expected, we observe an enhanced stress anisotropy in the tethers (i.e. the narrow regions between the circular holes), see Fig. \ref{patterned_membrane}a. 
In particular, a cut along a tether reveals a peak, symmetric with respect to the center of the tether, and a maximum stress anisotropy of $\Delta\sigma\sim 2.3~\mathrm{GPa}$. 
Excellent agreement with the predicted stress anisotropy in this region confirms the previously computed value of the Brewster coefficient.

Near the edges of the membrane, large retardation values ($\gtrsim 15~\mathrm{mrad}$) appears as dark blue rings around the holes (see Fig \ref{patterned_membrane}a).
We attribute these to diffraction-related imaging artifacts.
This is supported by the presence of several weak concentric rings around the holes, likely due to higher diffraction orders.
In addition, the observation of these features already in single-shot images rules out excentric rotation as their cause.

\begin{figure}[H]
\includegraphics[width = 0.5\textwidth]{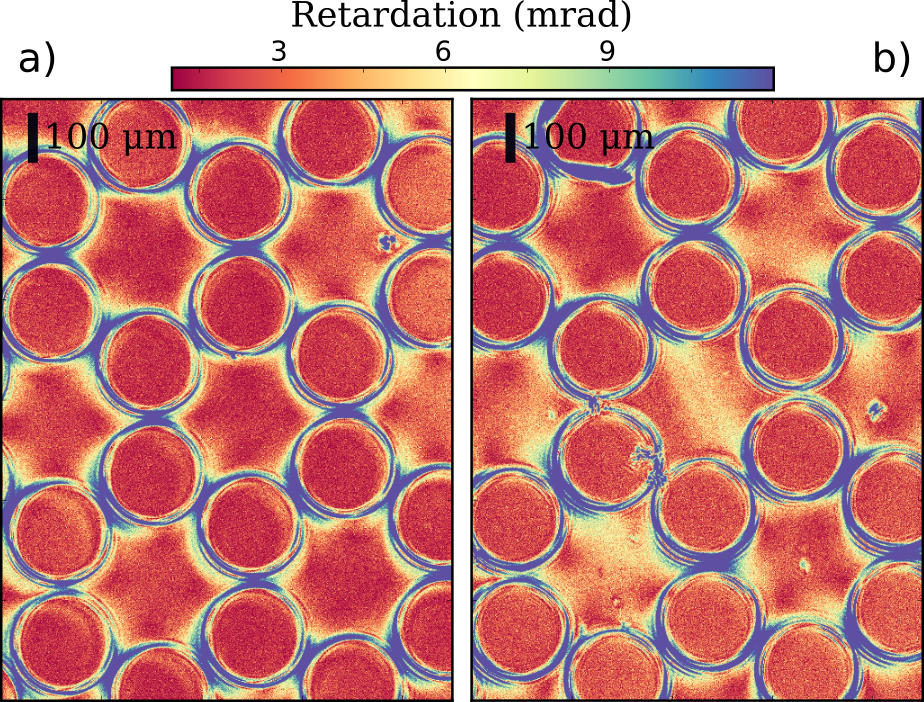}% Here is how to import EPS art
\caption{\label{broken_membrane} Comparison between an undamaged membrane (a) and one with a broken tether (b). Stress released in (b), and the broken periodicity can be clearly recognised, allowing identification and localisation of the defect.}
\end{figure}

In spite of such artifacts, it is straightforward to recognise defective membranes.
Figure \ref{broken_membrane} shows retardation images of an undamaged membrane and one with a broken tether, as a comparison.
Defects in the perforated membrane structure cause a redistribution of the stress. One could envision using this approach in assessing the overall performance of the device (e.g. the quality factor), by comparing the stress profile of the defective device with simulations of an undamaged membrane resonator.

In conclusion, applied to $210~\mathrm{nm}$-thick membrane resonators, our experimental setup achieves a $\sim 0.2~\,\mathrm{mrad}$ resolution in retardation over a $\sim(200~\mathrm{\mu m})^2$ area.
Using the measured Brewster coefficient, this corresponds to a resolution in stress anisotropy of $\sim 80~\mathrm{MPa}$.
Thus we expect the method to deliver relevant results also for ultrathin ($l\lesssim 50~\mathrm{nm}$) membranes of smaller dimensions.
As an imaging modality, the present, very basic microscopy setup achieves moderate transverse resolution on the order of $10~\mathrm{\mu m}$, impaired also by edge artifacts (likely diffraction-caused).
It is nonetheless more than sufficient to assess the overall quality of patterned membrane resonators and localise defects.
We have implemented a gray field polariscope optimised for extracting stress anisotropy in membrane resonators. 
We have validated the technique by comparing our measurement results with FEM simulations, and performed an independent measurement of the Brewster coefficient, correcting an older value found in the litterature\cite{Campillo2002}.
In itself, this result is of interest for optomechanical systems relying on photoelastic coupling in silicon nitride \cite{Borkje2012,Grutter2015}.

%\nocite{*}
\bibliographystyle{aipnum4-1}
\bibliography{cleaned_bibliography}% Produces the bibliography via BibTeX.

\begin{thebibliography}{30}%
\makeatletter
\providecommand \@ifxundefined [1]{%
 \@ifx{#1\undefined}
}%
\providecommand \@ifnum [1]{%
 \ifnum #1\expandafter \@firstoftwo
 \else \expandafter \@secondoftwo
 \fi
}%
\providecommand \@ifx [1]{%
 \ifx #1\expandafter \@firstoftwo
 \else \expandafter \@secondoftwo
 \fi
}%
\providecommand \natexlab [1]{#1}%
\providecommand \enquote  [1]{``#1''}%
\providecommand \bibnamefont  [1]{#1}%
\providecommand \bibfnamefont [1]{#1}%
\providecommand \citenamefont [1]{#1}%
\providecommand \href@noop [0]{\@secondoftwo}%
\providecommand \href [0]{\begingroup \@sanitize@url \@href}%
\providecommand \@href[1]{\@@startlink{#1}\@@href}%
\providecommand \@@href[1]{\endgroup#1\@@endlink}%
\providecommand \@sanitize@url [0]{\catcode `\\12\catcode `\$12\catcode
  `\&12\catcode `\#12\catcode `\^12\catcode `\_12\catcode `\%12\relax}%
\providecommand \@@startlink[1]{}%
\providecommand \@@endlink[0]{}%
\providecommand \url  [0]{\begingroup\@sanitize@url \@url }%
\providecommand \@url [1]{\endgroup\@href {#1}{\urlprefix }}%
\providecommand \urlprefix  [0]{URL }%
\providecommand \Eprint [0]{\href }%
\providecommand \doibase [0]{http://dx.doi.org/}%
\providecommand \selectlanguage [0]{\@gobble}%
\providecommand \bibinfo  [0]{\@secondoftwo}%
\providecommand \bibfield  [0]{\@secondoftwo}%
\providecommand \translation [1]{[#1]}%
\providecommand \BibitemOpen [0]{}%
\providecommand \bibitemStop [0]{}%
\providecommand \bibitemNoStop [0]{.\EOS\space}%
\providecommand \EOS [0]{\spacefactor3000\relax}%
\providecommand \BibitemShut  [1]{\csname bibitem#1\endcsname}%
\let\auto@bib@innerbib\@empty
%</preamble>
\bibitem [{\citenamefont {Zwickl}\ \emph {et~al.}(2008)\citenamefont {Zwickl},
  \citenamefont {Shanks}, \citenamefont {Jayich}, \citenamefont {Yang},
  \citenamefont {{Bleszynski Jayich}}, \citenamefont {Thompson},\ and\
  \citenamefont {Harris}}]{Zwickl2008}%
  \BibitemOpen
  \bibfield  {author} {\bibinfo {author} {\bibfnamefont {B.~M.}\ \bibnamefont
  {Zwickl}}, \bibinfo {author} {\bibfnamefont {W.~E.}\ \bibnamefont {Shanks}},
  \bibinfo {author} {\bibfnamefont {M.}~\bibnamefont {Jayich}}, \bibinfo
  {author} {\bibfnamefont {C.}~\bibnamefont {Yang}}, \bibinfo {author}
  {\bibfnamefont {C.}~\bibnamefont {{Bleszynski Jayich}}}, \bibinfo {author}
  {\bibfnamefont {J.~D.}\ \bibnamefont {Thompson}}, \ and\ \bibinfo {author}
  {\bibfnamefont {J.~G.~E.}\ \bibnamefont {Harris}},\ }\href {\doibase
  10.1063/1.2884191} {\bibfield  {journal} {\bibinfo  {journal} {Applied
  Physics Letters}\ }\textbf {\bibinfo {volume} {92}},\ \bibinfo {pages}
  {103125} (\bibinfo {year} {2008})}\BibitemShut {NoStop}%
\bibitem [{\citenamefont {Wilson}\ \emph {et~al.}(2009)\citenamefont {Wilson},
  \citenamefont {Regal}, \citenamefont {Papp},\ and\ \citenamefont
  {Kimble}}]{Wilson2009}%
  \BibitemOpen
  \bibfield  {author} {\bibinfo {author} {\bibfnamefont {D.~J.}\ \bibnamefont
  {Wilson}}, \bibinfo {author} {\bibfnamefont {C.~A.}\ \bibnamefont {Regal}},
  \bibinfo {author} {\bibfnamefont {S.~B.}\ \bibnamefont {Papp}}, \ and\
  \bibinfo {author} {\bibfnamefont {H.~J.}\ \bibnamefont {Kimble}},\
  }\href@noop {} {\bibfield  {journal} {\bibinfo  {journal} {Physical Review
  Letters}\ }\textbf {\bibinfo {volume} {103}},\ \bibinfo {pages} {207204}
  (\bibinfo {year} {2009})}\BibitemShut {NoStop}%
\bibitem [{\citenamefont {Purdy}\ \emph {et~al.}(2013)\citenamefont {Purdy},
  \citenamefont {Yu}, \citenamefont {Peterson}, \citenamefont {Kampel},\ and\
  \citenamefont {Regal}}]{Purdy2013}%
  \BibitemOpen
  \bibfield  {author} {\bibinfo {author} {\bibfnamefont {T.~P.}\ \bibnamefont
  {Purdy}}, \bibinfo {author} {\bibfnamefont {P.-L.}\ \bibnamefont {Yu}},
  \bibinfo {author} {\bibfnamefont {R.~W.}\ \bibnamefont {Peterson}}, \bibinfo
  {author} {\bibfnamefont {N.~S.}\ \bibnamefont {Kampel}}, \ and\ \bibinfo
  {author} {\bibfnamefont {C.~A.}\ \bibnamefont {Regal}},\ }\href {\doibase
  10.1103/PhysRevX.3.031012} {\bibfield  {journal} {\bibinfo  {journal}
  {Physical Review X}\ }\textbf {\bibinfo {volume} {3}},\ \bibinfo {pages}
  {031012} (\bibinfo {year} {2013})}\BibitemShut {NoStop}%
\bibitem [{\citenamefont {Purdy}, \citenamefont {Peterson},\ and\ \citenamefont
  {Regal}(2013)}]{Purdy2013a}%
  \BibitemOpen
  \bibfield  {author} {\bibinfo {author} {\bibfnamefont {T.~P.}\ \bibnamefont
  {Purdy}}, \bibinfo {author} {\bibfnamefont {R.~W.}\ \bibnamefont {Peterson}},
  \ and\ \bibinfo {author} {\bibfnamefont {C.}~\bibnamefont {Regal}},\
  }\href@noop {} {\ \textbf {\bibinfo {volume} {339}},\ \bibinfo {pages} {1}
  (\bibinfo {year} {2013})}\BibitemShut {NoStop}%
\bibitem [{\citenamefont {Zhou}\ \emph {et~al.}(2013)\citenamefont {Zhou},
  \citenamefont {Hocke}, \citenamefont {Schliesser}, \citenamefont {Marx},
  \citenamefont {Huebl}, \citenamefont {Gross},\ and\ \citenamefont
  {Kippenberg}}]{Zhou2013}%
  \BibitemOpen
  \bibfield  {author} {\bibinfo {author} {\bibfnamefont {X.}~\bibnamefont
  {Zhou}}, \bibinfo {author} {\bibfnamefont {F.}~\bibnamefont {Hocke}},
  \bibinfo {author} {\bibfnamefont {A.}~\bibnamefont {Schliesser}}, \bibinfo
  {author} {\bibfnamefont {A.}~\bibnamefont {Marx}}, \bibinfo {author}
  {\bibfnamefont {H.~.}\ \bibnamefont {Huebl}}, \bibinfo {author}
  {\bibfnamefont {R.}~\bibnamefont {Gross}}, \ and\ \bibinfo {author}
  {\bibfnamefont {T.~J.}\ \bibnamefont {Kippenberg}},\ }\href {\doibase
  10.1038/nphys2527} {\bibfield  {journal} {\bibinfo  {journal} {Nature
  Physics}\ }\textbf {\bibinfo {volume} {9}},\ \bibinfo {pages} {179} (\bibinfo
  {year} {2013})}\BibitemShut {NoStop}%
\bibitem [{\citenamefont {Underwood}\ \emph {et~al.}(2015)\citenamefont
  {Underwood}, \citenamefont {Mason}, \citenamefont {Lee}, \citenamefont {Xu},
  \citenamefont {Jiang}, \citenamefont {Shkarin}, \citenamefont {Borkje},
  \citenamefont {Girvin},\ and\ \citenamefont {Harris}}]{Underwood2015}%
  \BibitemOpen
  \bibfield  {author} {\bibinfo {author} {\bibfnamefont {M.}~\bibnamefont
  {Underwood}}, \bibinfo {author} {\bibfnamefont {D.}~\bibnamefont {Mason}},
  \bibinfo {author} {\bibfnamefont {D.}~\bibnamefont {Lee}}, \bibinfo {author}
  {\bibfnamefont {H.}~\bibnamefont {Xu}}, \bibinfo {author} {\bibfnamefont
  {L.}~\bibnamefont {Jiang}}, \bibinfo {author} {\bibfnamefont {A.~B.}\
  \bibnamefont {Shkarin}}, \bibinfo {author} {\bibfnamefont {K.}~\bibnamefont
  {Borkje}}, \bibinfo {author} {\bibfnamefont {S.~M.}\ \bibnamefont {Girvin}},
  \ and\ \bibinfo {author} {\bibfnamefont {J.~G.~E.}\ \bibnamefont {Harris}},\
  }\href {\doibase 10.1103/physreva.92.061801} {\bibfield  {journal} {\bibinfo
  {journal} {Physical Review A}\ }\textbf {\bibinfo {volume} {92}},\ \bibinfo
  {pages} {061801(R)} (\bibinfo {year} {2015})}\BibitemShut {NoStop}%
\bibitem [{\citenamefont {Wilson}\ \emph {et~al.}(2015)\citenamefont {Wilson},
  \citenamefont {Sudhir}, \citenamefont {Piro}, \citenamefont {Schilling},
  \citenamefont {Ghadimi},\ and\ \citenamefont {Kippenberg}}]{Wilson2015}%
  \BibitemOpen
  \bibfield  {author} {\bibinfo {author} {\bibfnamefont {D.~J.}\ \bibnamefont
  {Wilson}}, \bibinfo {author} {\bibfnamefont {V.}~\bibnamefont {Sudhir}},
  \bibinfo {author} {\bibfnamefont {N.}~\bibnamefont {Piro}}, \bibinfo {author}
  {\bibfnamefont {R.}~\bibnamefont {Schilling}}, \bibinfo {author}
  {\bibfnamefont {A.}~\bibnamefont {Ghadimi}}, \ and\ \bibinfo {author}
  {\bibfnamefont {T.~J.}\ \bibnamefont {Kippenberg}},\ }\href {\doibase
  10.1038/nature14672} {\bibfield  {journal} {\bibinfo  {journal} {Nature}\
  }\textbf {\bibinfo {volume} {524}},\ \bibinfo {pages} {325} (\bibinfo {year}
  {2015})}\BibitemShut {NoStop}%
\bibitem [{\citenamefont {Peterson}\ \emph {et~al.}(2016)\citenamefont
  {Peterson}, \citenamefont {Purdy}, \citenamefont {Kampel}, \citenamefont
  {Andrews}, \citenamefont {Yu}, \citenamefont {Lehnert},\ and\ \citenamefont
  {Regal}}]{Peterson2016}%
  \BibitemOpen
  \bibfield  {author} {\bibinfo {author} {\bibfnamefont {R.}~\bibnamefont
  {Peterson}}, \bibinfo {author} {\bibfnamefont {T.~P.}\ \bibnamefont {Purdy}},
  \bibinfo {author} {\bibfnamefont {N.}~\bibnamefont {Kampel}}, \bibinfo
  {author} {\bibfnamefont {R.}~\bibnamefont {Andrews}}, \bibinfo {author}
  {\bibfnamefont {P.-L.}\ \bibnamefont {Yu}}, \bibinfo {author} {\bibfnamefont
  {K.}~\bibnamefont {Lehnert}}, \ and\ \bibinfo {author} {\bibfnamefont
  {C.}~\bibnamefont {Regal}},\ }\href {\doibase 10.1103/PhysRevLett.116.063601}
  {\bibfield  {journal} {\bibinfo  {journal} {Physical Review Letters}\
  }\textbf {\bibinfo {volume} {116}},\ \bibinfo {pages} {063601} (\bibinfo
  {year} {2016})},\ \Eprint {http://arxiv.org/abs/1510.03911} {1510.03911}
  \BibitemShut {NoStop}%
\bibitem [{\citenamefont {Nielsen}\ \emph {et~al.}(2016)\citenamefont
  {Nielsen}, \citenamefont {Tsaturyan}, \citenamefont {M{\o}ller},
  \citenamefont {Polzik},\ and\ \citenamefont {Schliesser}}]{Nielsen2016}%
  \BibitemOpen
  \bibfield  {author} {\bibinfo {author} {\bibfnamefont {W.~H.~P.}\
  \bibnamefont {Nielsen}}, \bibinfo {author} {\bibfnamefont {Y.}~\bibnamefont
  {Tsaturyan}}, \bibinfo {author} {\bibfnamefont {C.~B.}\ \bibnamefont
  {M{\o}ller}}, \bibinfo {author} {\bibfnamefont {E.~S.}\ \bibnamefont
  {Polzik}}, \ and\ \bibinfo {author} {\bibfnamefont {A.}~\bibnamefont
  {Schliesser}},\ }\href {\doibase 10.1073/pnas.1608412114} {\bibfield
  {journal} {\bibinfo  {journal} {PNAS}\ }\textbf {\bibinfo {volume} {114}},\
  \bibinfo {pages} {62–} (\bibinfo {year} {2016})}\BibitemShut {NoStop}%
\bibitem [{\citenamefont {Noguchi}\ \emph {et~al.}(2016)\citenamefont
  {Noguchi}, \citenamefont {Yamazaki}, \citenamefont {Ataka}, \citenamefont
  {Fujita}, \citenamefont {Tabuchi}, \citenamefont {Ishikawa}, \citenamefont
  {Usami},\ and\ \citenamefont {Nakamura}}]{Koji2016}%
  \BibitemOpen
  \bibfield  {author} {\bibinfo {author} {\bibfnamefont {A.}~\bibnamefont
  {Noguchi}}, \bibinfo {author} {\bibfnamefont {R.}~\bibnamefont {Yamazaki}},
  \bibinfo {author} {\bibfnamefont {M.}~\bibnamefont {Ataka}}, \bibinfo
  {author} {\bibfnamefont {H.}~\bibnamefont {Fujita}}, \bibinfo {author}
  {\bibfnamefont {Y.}~\bibnamefont {Tabuchi}}, \bibinfo {author} {\bibfnamefont
  {T.}~\bibnamefont {Ishikawa}}, \bibinfo {author} {\bibfnamefont
  {K.}~\bibnamefont {Usami}}, \ and\ \bibinfo {author} {\bibfnamefont
  {Y.}~\bibnamefont {Nakamura}},\ }\href@noop {} {\bibfield  {journal}
  {\bibinfo  {journal} {New journal of Physics}\ }\textbf {\bibinfo {volume}
  {18}} (\bibinfo {year} {2016})}\BibitemShut {NoStop}%
\bibitem [{\citenamefont {Stambaugh}\ \emph {et~al.}(2015)\citenamefont
  {Stambaugh}, \citenamefont {Xu}, \citenamefont {Kemiktarak}, \citenamefont
  {Taylor},\ and\ \citenamefont {Lawall}}]{Stambaugh2014}%
  \BibitemOpen
  \bibfield  {author} {\bibinfo {author} {\bibfnamefont {C.}~\bibnamefont
  {Stambaugh}}, \bibinfo {author} {\bibfnamefont {H.}~\bibnamefont {Xu}},
  \bibinfo {author} {\bibfnamefont {U.}~\bibnamefont {Kemiktarak}}, \bibinfo
  {author} {\bibfnamefont {J.}~\bibnamefont {Taylor}}, \ and\ \bibinfo {author}
  {\bibfnamefont {J.}~\bibnamefont {Lawall}},\ }\href@noop {} {\bibfield
  {journal} {\bibinfo  {journal} {Annalen der Physik}\ }\textbf {\bibinfo
  {volume} {527}},\ \bibinfo {pages} {81} (\bibinfo {year} {2015})},\ \Eprint
  {http://arxiv.org/abs/1407.1709} {1407.1709} \BibitemShut {NoStop}%
\bibitem [{\citenamefont {Chen}\ \emph {et~al.}(2016)\citenamefont {Chen},
  \citenamefont {Chardin}, \citenamefont {Makles}, \citenamefont {Ca{\"er}},
  \citenamefont {Chua}, \citenamefont {Braive}, \citenamefont {Robert-Philip},
  \citenamefont {Briant}, \citenamefont {Cohadon}, \citenamefont {Heidmann},
  \citenamefont {Jacqmin},\ and\ \citenamefont {Del{\'e}glise}}]{Chen2016a}%
  \BibitemOpen
  \bibfield  {author} {\bibinfo {author} {\bibfnamefont {X.}~\bibnamefont
  {Chen}}, \bibinfo {author} {\bibfnamefont {C.}~\bibnamefont {Chardin}},
  \bibinfo {author} {\bibfnamefont {K.}~\bibnamefont {Makles}}, \bibinfo
  {author} {\bibfnamefont {C.}~\bibnamefont {Ca{\"er}}}, \bibinfo {author}
  {\bibfnamefont {S.}~\bibnamefont {Chua}}, \bibinfo {author} {\bibfnamefont
  {R.}~\bibnamefont {Braive}}, \bibinfo {author} {\bibfnamefont
  {I.}~\bibnamefont {Robert-Philip}}, \bibinfo {author} {\bibfnamefont
  {T.}~\bibnamefont {Briant}}, \bibinfo {author} {\bibfnamefont {P.-F.}\
  \bibnamefont {Cohadon}}, \bibinfo {author} {\bibfnamefont {A.}~\bibnamefont
  {Heidmann}}, \bibinfo {author} {\bibfnamefont {T.}~\bibnamefont {Jacqmin}}, \
  and\ \bibinfo {author} {\bibfnamefont {S.}~\bibnamefont {Del{\'e}glise}},\
  }\href@noop {} {\bibfield  {journal} {\bibinfo  {journal} {arXiv:1603.07200}\
  } (\bibinfo {year} {2016})}\BibitemShut {NoStop}%
\bibitem [{\citenamefont {Norte}, \citenamefont {Moura},\ and\ \citenamefont
  {Gr{\"o}blacher}(2016)}]{Norte2016}%
  \BibitemOpen
  \bibfield  {author} {\bibinfo {author} {\bibfnamefont {R.}~\bibnamefont
  {Norte}}, \bibinfo {author} {\bibfnamefont {J.~P.}\ \bibnamefont {Moura}}, \
  and\ \bibinfo {author} {\bibfnamefont {S.}~\bibnamefont {Gr{\"o}blacher}},\
  }\href {\doibase 10.1103/physrevlett.116.147202} {\bibfield  {journal}
  {\bibinfo  {journal} {Physical Review Letters}\ }\textbf {\bibinfo {volume}
  {116}},\ \bibinfo {pages} {147202} (\bibinfo {year} {2016})}\BibitemShut
  {NoStop}%
\bibitem [{\citenamefont {Bernard}\ \emph {et~al.}(2016)\citenamefont
  {Bernard}, \citenamefont {Reinhardt}, \citenamefont {Dumont}, \citenamefont
  {Peter},\ and\ \citenamefont {Sankey}}]{Bernard2016}%
  \BibitemOpen
  \bibfield  {author} {\bibinfo {author} {\bibfnamefont {S.}~\bibnamefont
  {Bernard}}, \bibinfo {author} {\bibfnamefont {C.}~\bibnamefont {Reinhardt}},
  \bibinfo {author} {\bibfnamefont {V.}~\bibnamefont {Dumont}}, \bibinfo
  {author} {\bibfnamefont {Y.-A.}\ \bibnamefont {Peter}}, \ and\ \bibinfo
  {author} {\bibfnamefont {J.~C.}\ \bibnamefont {Sankey}},\ }\href@noop {}
  {\bibfield  {journal} {\bibinfo  {journal} {Optics Letter}\ }\textbf
  {\bibinfo {volume} {41}},\ \bibinfo {pages} {5624} (\bibinfo {year}
  {2016})}\BibitemShut {NoStop}%
\bibitem [{\citenamefont {Kleckner}\ \emph {et~al.}(2011)\citenamefont
  {Kleckner}, \citenamefont {Pepper}, \citenamefont {Jeffrey}, \citenamefont
  {Sonin}, \citenamefont {Thon},\ and\ \citenamefont
  {Bouwmeester}}]{Kleckner2011}%
  \BibitemOpen
  \bibfield  {author} {\bibinfo {author} {\bibfnamefont {D.}~\bibnamefont
  {Kleckner}}, \bibinfo {author} {\bibfnamefont {B.}~\bibnamefont {Pepper}},
  \bibinfo {author} {\bibfnamefont {E.}~\bibnamefont {Jeffrey}}, \bibinfo
  {author} {\bibfnamefont {P.}~\bibnamefont {Sonin}}, \bibinfo {author}
  {\bibfnamefont {S.~M.}\ \bibnamefont {Thon}}, \ and\ \bibinfo {author}
  {\bibfnamefont {D.}~\bibnamefont {Bouwmeester}},\ }\href {\doibase
  10.1364/oe.19.019708} {\bibfield  {journal} {\bibinfo  {journal} {Optics
  Express}\ }\textbf {\bibinfo {volume} {19}},\ \bibinfo {pages} {19708}
  (\bibinfo {year} {2011})}\BibitemShut {NoStop}%
\bibitem [{\citenamefont {Reinhardt}\ \emph {et~al.}(2016)\citenamefont
  {Reinhardt}, \citenamefont {M{\"u}ller}, \citenamefont {Bourassa},\ and\
  \citenamefont {Sankey}}]{Reinhardt2016}%
  \BibitemOpen
  \bibfield  {author} {\bibinfo {author} {\bibfnamefont {C.}~\bibnamefont
  {Reinhardt}}, \bibinfo {author} {\bibfnamefont {T.}~\bibnamefont
  {M{\"u}ller}}, \bibinfo {author} {\bibfnamefont {A.}~\bibnamefont
  {Bourassa}}, \ and\ \bibinfo {author} {\bibfnamefont {J.~C.}\ \bibnamefont
  {Sankey}},\ }\href {\doibase 10.1103/physrevx.6.021001} {\bibfield  {journal}
  {\bibinfo  {journal} {Physical Review X}\ }\textbf {\bibinfo {volume} {6}},\
  \bibinfo {pages} {021001} (\bibinfo {year} {2016})}\BibitemShut {NoStop}%
\bibitem [{\citenamefont {Weaver}\ \emph {et~al.}(2016)\citenamefont {Weaver},
  \citenamefont {Pepper}, \citenamefont {Luna}, \citenamefont {Buters},
  \citenamefont {Eerkens}, \citenamefont {Welker}, \citenamefont {Perock},
  \citenamefont {Heeck}, \citenamefont {de~Man},\ and\ \citenamefont
  {Bouwmeester}}]{Weaver2016}%
  \BibitemOpen
  \bibfield  {author} {\bibinfo {author} {\bibfnamefont {M.~J.}\ \bibnamefont
  {Weaver}}, \bibinfo {author} {\bibfnamefont {B.}~\bibnamefont {Pepper}},
  \bibinfo {author} {\bibfnamefont {F.}~\bibnamefont {Luna}}, \bibinfo {author}
  {\bibfnamefont {F.~M.}\ \bibnamefont {Buters}}, \bibinfo {author}
  {\bibfnamefont {H.~J.}\ \bibnamefont {Eerkens}}, \bibinfo {author}
  {\bibfnamefont {G.}~\bibnamefont {Welker}}, \bibinfo {author} {\bibfnamefont
  {B.}~\bibnamefont {Perock}}, \bibinfo {author} {\bibfnamefont
  {K.}~\bibnamefont {Heeck}}, \bibinfo {author} {\bibfnamefont
  {S.}~\bibnamefont {de~Man}}, \ and\ \bibinfo {author} {\bibfnamefont
  {D.}~\bibnamefont {Bouwmeester}},\ }\href {\doibase 10.1063/1.4939828}
  {\bibfield  {journal} {\bibinfo  {journal} {Applied Physics Letters}\
  }\textbf {\bibinfo {volume} {108}},\ \bibinfo {pages} {033501} (\bibinfo
  {year} {2016})}\BibitemShut {NoStop}%
\bibitem [{\citenamefont {Ghadimi}, \citenamefont {Wilson},\ and\ \citenamefont
  {Kippenberg}(2016)}]{Ghadimi2016}%
  \BibitemOpen
  \bibfield  {author} {\bibinfo {author} {\bibfnamefont {A.~H.}\ \bibnamefont
  {Ghadimi}}, \bibinfo {author} {\bibfnamefont {D.~J.}\ \bibnamefont {Wilson}},
  \ and\ \bibinfo {author} {\bibfnamefont {T.~J.}\ \bibnamefont {Kippenberg}},\
  }\href@noop {} {\bibfield  {journal} {\bibinfo  {journal} {arXiv:1603.01605}\
  } (\bibinfo {year} {2016})}\BibitemShut {NoStop}%
\bibitem [{\citenamefont {Barasheed}, \citenamefont {M{\"u}ller},\ and\
  \citenamefont {Sankey}(2015)}]{Barasheed2015}%
  \BibitemOpen
  \bibfield  {author} {\bibinfo {author} {\bibfnamefont {A.~Z.}\ \bibnamefont
  {Barasheed}}, \bibinfo {author} {\bibfnamefont {T.}~\bibnamefont
  {M{\"u}ller}}, \ and\ \bibinfo {author} {\bibfnamefont {J.~C.}\ \bibnamefont
  {Sankey}},\ }\href@noop {} {\bibfield  {journal} {\bibinfo  {journal}
  {arXiv:1511.06193}\ } (\bibinfo {year} {2015})}\BibitemShut {NoStop}%
\bibitem [{\citenamefont {Tsaturyan}\ \emph {et~al.}(2016)\citenamefont
  {Tsaturyan}, \citenamefont {Barg}, \citenamefont {Polzik},\ and\
  \citenamefont {Schliesser}}]{Tsaturyan2016}%
  \BibitemOpen
  \bibfield  {author} {\bibinfo {author} {\bibfnamefont {Y.}~\bibnamefont
  {Tsaturyan}}, \bibinfo {author} {\bibfnamefont {A.}~\bibnamefont {Barg}},
  \bibinfo {author} {\bibfnamefont {E.~S.}\ \bibnamefont {Polzik}}, \ and\
  \bibinfo {author} {\bibfnamefont {A.}~\bibnamefont {Schliesser}},\
  }\href@noop {} {\bibfield  {journal} {\bibinfo  {journal} {arXiv:1603.07200}\
  } (\bibinfo {year} {2016})}\BibitemShut {NoStop}%
\bibitem [{\citenamefont {B{\o}rkje}\ and\ \citenamefont
  {Girvin}(2012)}]{Borkje2012}%
  \BibitemOpen
  \bibfield  {author} {\bibinfo {author} {\bibfnamefont {K.}~\bibnamefont
  {B{\o}rkje}}\ and\ \bibinfo {author} {\bibfnamefont {S.~M.}\ \bibnamefont
  {Girvin}},\ }\href {\doibase 10.1088/1367-2630/14/8/085016} {\bibfield
  {journal} {\bibinfo  {journal} {New J. Phys.}\ }\textbf {\bibinfo {volume}
  {14}},\ \bibinfo {pages} {085016} (\bibinfo {year} {2012})}\BibitemShut
  {NoStop}%
\bibitem [{\citenamefont {Grutter}, \citenamefont {Davanco},\ and\
  \citenamefont {Srinivasan}(2015)}]{Grutter2015}%
  \BibitemOpen
  \bibfield  {author} {\bibinfo {author} {\bibfnamefont {K.~E.}\ \bibnamefont
  {Grutter}}, \bibinfo {author} {\bibfnamefont {M.}~\bibnamefont {Davanco}}, \
  and\ \bibinfo {author} {\bibfnamefont {K.}~\bibnamefont {Srinivasan}},\
  }\href {\doibase 10.1109/jstqe.2014.2376966} {\bibfield  {journal} {\bibinfo
  {journal} {IEEE Journal of Selected Topics in Quantum Electronics}\ }\textbf
  {\bibinfo {volume} {21}},\ \bibinfo {pages} {61} (\bibinfo {year}
  {2015})}\BibitemShut {NoStop}%
\bibitem [{\citenamefont {Ajovalasit}, \citenamefont {Petrucci},\ and\
  \citenamefont {Scafidi}(2015)}]{Ajovalasit2015}%
  \BibitemOpen
  \bibfield  {author} {\bibinfo {author} {\bibfnamefont {A.}~\bibnamefont
  {Ajovalasit}}, \bibinfo {author} {\bibfnamefont {G.}~\bibnamefont
  {Petrucci}}, \ and\ \bibinfo {author} {\bibfnamefont {M.}~\bibnamefont
  {Scafidi}},\ }\href {\doibase 10.1111/ext.12017} {\bibfield  {journal}
  {\bibinfo  {journal} {Experimental Techniques}\ }\textbf {\bibinfo {volume}
  {39}},\ \bibinfo {pages} {11} (\bibinfo {year} {2015})}\BibitemShut {NoStop}%
\bibitem [{Note1()}]{Note1}%
  \BibitemOpen
  \bibinfo {note} {This measurement was conducted on a separate sample with a
  $238.6$-nm thick SiN layer grown in the same conditions as the
  ribbon.}\BibitemShut {Stop}%
\bibitem [{Note2()}]{Note2}%
  \BibitemOpen
  \bibinfo {note} {The transmittance of a membrane of thickness $d$ and
  refractive index $n$ is given by $t = \protect \frac
  {2ine^{-ikd}}{2in\protect \qopname \relax o{cos}(kdn)+(1+n^2)\protect
  \qopname \relax o{sin}(nkd)}$}\BibitemShut {NoStop}%
\bibitem [{\citenamefont {Campillo}\ and\ \citenamefont
  {Hsu}(2002)}]{Campillo2002}%
  \BibitemOpen
  \bibfield  {author} {\bibinfo {author} {\bibfnamefont {A.~L.}\ \bibnamefont
  {Campillo}}\ and\ \bibinfo {author} {\bibfnamefont {J.~W.~P.}\ \bibnamefont
  {Hsu}},\ }\href {\doibase 10.1063/1.1415065} {\bibfield  {journal} {\bibinfo
  {journal} {Journal of Applied Physics}\ }\textbf {\bibinfo {volume} {91}},\
  \bibinfo {pages} {646} (\bibinfo {year} {2002})}\BibitemShut {NoStop}%
\bibitem [{\citenamefont {Priestley}(2001)}]{Priestley}%
  \BibitemOpen
  \bibfield  {author} {\bibinfo {author} {\bibfnamefont {R.}~\bibnamefont
  {Priestley}},\ }\href@noop {} {\bibfield  {journal} {\bibinfo  {journal}
  {Proceedings of SPIE}\ }\textbf {\bibinfo {volume} {4346}} (\bibinfo {year}
  {2001})}\BibitemShut {NoStop}%
\bibitem [{\citenamefont {Tsaturyan}\ \emph {et~al.}(2014)\citenamefont
  {Tsaturyan}, \citenamefont {Barg}, \citenamefont {Simonsen}, \citenamefont
  {Villanueva}, \citenamefont {Schmid}, \citenamefont {Schliesser},\ and\
  \citenamefont {Polzik}}]{Tsaturyan2014}%
  \BibitemOpen
  \bibfield  {author} {\bibinfo {author} {\bibfnamefont {Y.}~\bibnamefont
  {Tsaturyan}}, \bibinfo {author} {\bibfnamefont {A.}~\bibnamefont {Barg}},
  \bibinfo {author} {\bibfnamefont {A.}~\bibnamefont {Simonsen}}, \bibinfo
  {author} {\bibfnamefont {L.~G.}\ \bibnamefont {Villanueva}}, \bibinfo
  {author} {\bibfnamefont {S.}~\bibnamefont {Schmid}}, \bibinfo {author}
  {\bibfnamefont {A.}~\bibnamefont {Schliesser}}, \ and\ \bibinfo {author}
  {\bibfnamefont {E.~S.}\ \bibnamefont {Polzik}},\ }\href {\doibase
  10.1364/OE.22.006810} {\bibfield  {journal} {\bibinfo  {journal} {Optics
  Express}\ }\textbf {\bibinfo {volume} {22}},\ \bibinfo {pages} {6810}
  (\bibinfo {year} {2014})}\BibitemShut {NoStop}%
\bibitem [{\citenamefont {Yu}\ \emph {et~al.}(2014)\citenamefont {Yu},
  \citenamefont {Cicak}, \citenamefont {Kampel}, \citenamefont {Tsaturyan},
  \citenamefont {Purdy}, \citenamefont {Simmonds},\ and\ \citenamefont
  {Regal}}]{Yu2014}%
  \BibitemOpen
  \bibfield  {author} {\bibinfo {author} {\bibfnamefont {P.~L.}\ \bibnamefont
  {Yu}}, \bibinfo {author} {\bibfnamefont {K.}~\bibnamefont {Cicak}}, \bibinfo
  {author} {\bibfnamefont {N.~S.}\ \bibnamefont {Kampel}}, \bibinfo {author}
  {\bibfnamefont {Y.}~\bibnamefont {Tsaturyan}}, \bibinfo {author}
  {\bibfnamefont {T.~P.}\ \bibnamefont {Purdy}}, \bibinfo {author}
  {\bibfnamefont {R.~W.}\ \bibnamefont {Simmonds}}, \ and\ \bibinfo {author}
  {\bibfnamefont {C.~A.}\ \bibnamefont {Regal}},\ }\href {\doibase
  10.1063/1.4862031} {\bibfield  {journal} {\bibinfo  {journal} {Applied
  Physics Letters}\ }\textbf {\bibinfo {volume} {104}} (\bibinfo {year}
  {2014}),\ 10.1063/1.4862031}\BibitemShut {NoStop}%
\bibitem [{\citenamefont {Wilson-Rae}(2008)}]{Wilson-Rae2008}%
  \BibitemOpen
  \bibfield  {author} {\bibinfo {author} {\bibfnamefont {I.}~\bibnamefont
  {Wilson-Rae}},\ }\href {\doibase 10.1103/PhysRevB.77.245418} {\bibfield
  {journal} {\bibinfo  {journal} {Physical Review B - Condensed Matter and
  Materials Physics}\ }\textbf {\bibinfo {volume} {77}} (\bibinfo {year}
  {2008}),\ 10.1103/PhysRevB.77.245418},\ \Eprint
  {http://arxiv.org/abs/0710.0200} {arXiv:0710.0200} \BibitemShut {NoStop}%
\end{thebibliography}%

\end{document}